\documentclass[9pt,twocolumn]{extarticle}

\usepackage{graphicx}
\usepackage{amssymb}
\usepackage[english]{babel}
\usepackage{color}
\usepackage{cite}

\begin{document}

\title{\bf Membrane tubulation by elongated and patchy nanoparticles}

\author{Michael Raatz\footnote{Present address: Department of Ecology and Ecosystem Modeling, Institute for Biochemistry and Biology, University of Potsdam, Am Neuen Palais 10, 14469 Potsdam, Germany}  ~and Thomas R.\ Weikl \\ \small Max Planck Institute of Colloids and Interfaces, Department of Theory and Bio-Systems, 14424 Potsdam, Germany}

\date{}

\maketitle

\begin{abstract}
Advances in nanotechnology lead to an increasing interest in how nanoparticles interact with biomembranes. Nanoparticles are wrapped spontaneously by biomembranes if the adhesive interactions between the particles and membranes compensate for the cost of membrane bending. In the last years, the cooperative wrapping of spherical nanoparticles in membrane tubules has been observed in experiments and simulations. For spherical nanoparticles, the stability of the particle-filled membrane tubules strongly depends on the range of the adhesive particle-membrane interactions. In this article, we show via modeling and energy minimization that elongated and patchy particles are wrapped cooperatively in membrane tubules that are highly stable for all ranges of the particle-membrane interactions, compared to individual wrapping of the particles. The cooperative wrapping of linear chains of elongated or patchy particles in membrane tubules may thus provide an efficient route to induce membrane tubulation, or to store such particles in membranes. 
\end{abstract}
%

%%%
\section{Introduction}
%%%

Because of their fluidity and flexilibity, biomembranes can adopt a variety of morphologies \cite{Zimmerberg06,Dimova06}.  Among these morphologies are membrane tubules with typical diameters of tens of nanometers. In biological cells, membrane tubules are stabilized by the adsorption of scaffolding proteins \cite{Takei99,Peter04,Voeltz06,Frost08,Shibata10}. In reconstituted lipid membrane systems, membrane tubules have been induced by antimicrobial peptides \cite{Domanov06,Mally07,Arouri11}, by aqueous phase separation inside vesicles \cite{Li11,Lipowsky13}, and by nanoparticles \cite{Yu09,Gozen11}. Tubules that tightly wrap linear aggregates of simian virus 40 particles have been found to occur both in cellular and reconstituted membranes \cite{Ewers10}. Similar membrane tubules filled with spherical nanoparticles have been observed by several groups in simulations \cite{Bahrami12,Saric12b,Yue12}. For spherical nanoparticles, the stability of these tubules strongly depends on the range of the adhesive particle-membrane interaction that causes the tubulation \cite{Raatz14}.

\begin{figure}[htp]
\begin{center}
\resizebox{0.6\columnwidth}{!}{\includegraphics{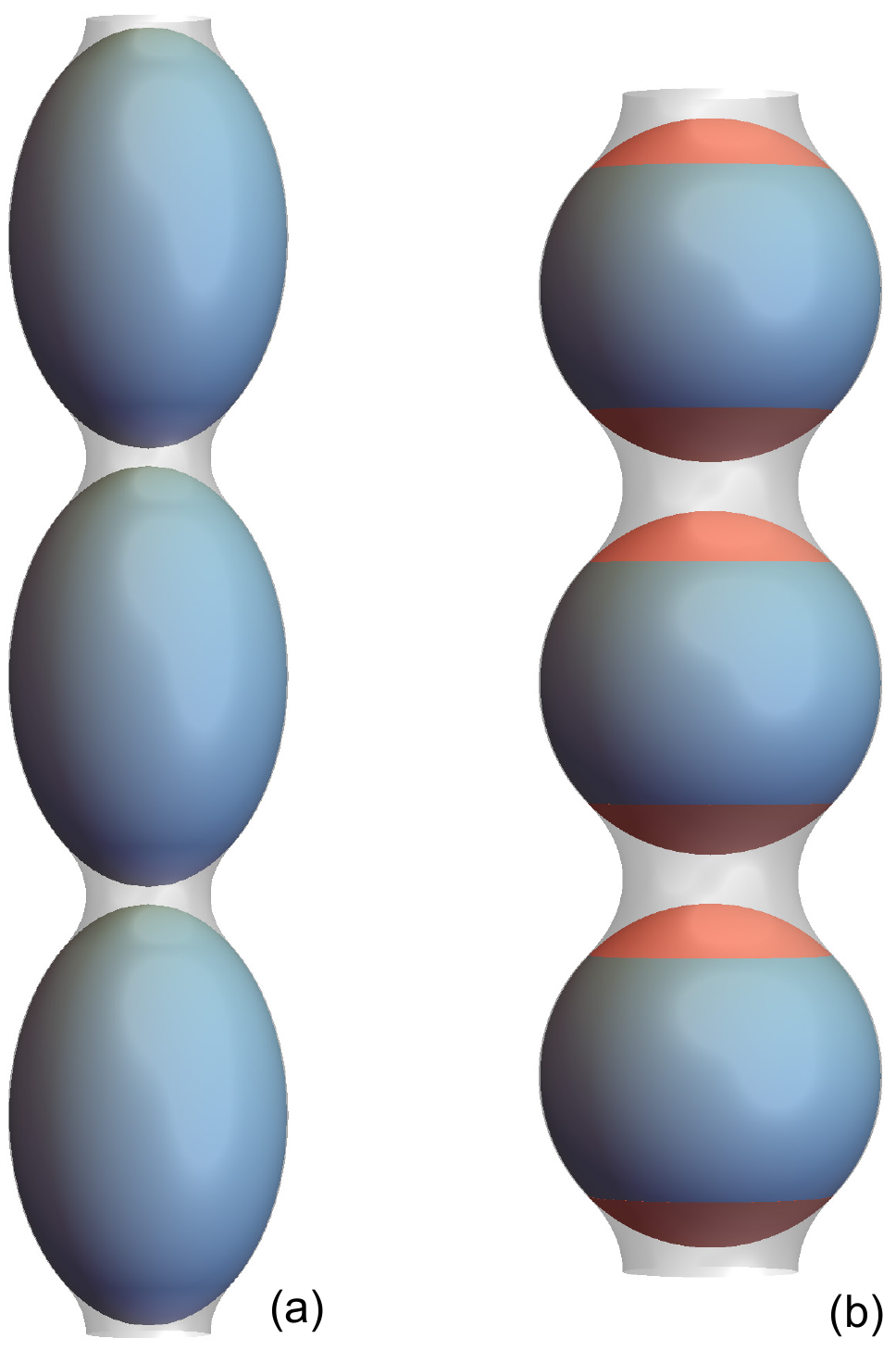}}
\end{center}
\caption{Minimum-energy shapes of membrane tubules around (a) prolate particles with aspect ratio $r = 1.5$ and (b) triblock Janus particles with strongly adhesive central  surface segment (blue) and non-adhesive tips (red). For prolate particles, the interplay of the bending and adhesion energies during membrane wrapping can be characterized by the rescaled adhesion energy $u = U R^2/\kappa$ where $U$ is the adhesion energy per area, $\kappa$ is the bending rigidity of the membranes, and $R$ is the radius of a sphere with the same surface area as the prolate particle. For the rescaled adhesion energy $u = 3$, the more strongly curved tips of the prolate particles in (a) remain unwrapped. The range of the particle-membrane interaction here is taken to be negligibly small compared to the particle dimensions. 
}
\label{figure-3D}
\end{figure}

In this article, we show that elongated and patchy nanoparticles can induce particle-filled tubules that are highly stable for all ranges of the particle-membrane interaction. These membrane tubules arise from the interplay of the adhesion energy of the particles and the bending energy of the membranes. Fig.\ \ref{figure-3D} illustrates segments of tubules filled (a) with prolate particles and (b) with patchy triblock Janus particles. Prolate particles have a rather high mean curvature at their tips and a lower mean curvature at their sides. The adhesion of membranes to the  tips of prolates therefore costs more bending energy than adhesion to the sides. The cooperative wrapping of the prolate particles in the membrane tubule of Fig.\ \ref{figure-3D}(a) is energetically favorable compared to the individual wrapping of the particles because the tubular membrane only adheres to the sides of the prolates and not to the tips. The individual wrapping of the prolate particles, in contrast, requires membrane adhesion at one of the tips, which costs additional bending energy. The non-adhering membrane regions of the tubules in between neighboring particles adopt a catenoidal shape of zero mean curvature and, thus, zero bending energy. The triblock Janus particles of Fig.\ \ref{figure-3D}(b) have non-adhesive tips (red) and strongly adhesive sides (blue), which are covered by the membrane in the tubule. 

How nanoparticles interact with biomembranes has been investigated with great intensity in the last years. The focus of these investigations has been largely on the wrapping of individual nanoparticles by biomembranes, and on how this wrapping is affected by the size and shape of the nanoparticles.  The wrapping of nanoparticles by membranes can either occur spontaneously from an interplay of bending and adhesion energies, or can be assisted by the curvature-inducing proteins and protein machineries of cellular membranes \cite{Mukherjee97,Conner03,Hurley10,Rodriguez13}. The spontaneous wapping of individual nanoparticles has been observed in experiments with lipid vesicles \cite{Dietrich97,Koltover99,Fery03,LeBihan09,Michel12}, polymersomes \cite{Jaskiewicz12,Jaskiewicz12b}, and cells \cite{Rothen06,Liu11}, and has been investigated in theoretical approaches\cite{Lipowsky98,Deserno02,Deserno04,Fleck04,Gozdz07,Benoit07,Nowak08,Decuzzi08,Chen09,Yi11,Cao11,Raatz14,Agudo15,Bahrami16}  and simulations  \cite{Noguchi02,Smith07,Fosnaric09,Li10,Yang10,Vacha11,Shi11,Yue11,Bahrami12,Saric12b,Yue12,Li12,Ding12,Dasgupta13,Bahrami13,Dasgupta14,Huang13,Curtis15,Ding15}.  

The cooperative wrapping of prolate or triblock Janus particles in membrane tubules provides a route to induce membrane tubulation. The membrane tubules filled with linear chains of such particles are highly stable relative to the individual wrapping of these particles. Membrane tubules also allow to encapsulate nanoparticles reversibly in vesicle membranes.   The amount of encapsulated nanoparticles and their release can be controlled by adjusting the area-to-volume ratio of the vesicles, i.e.\ by deflation and inflation of the vesicles induced by changes in osmotic conditions \cite{Bahrami13}.

%%%
\section{Model}
%%%

The spontaneous wrapping of nanoparticles by biomembranes results from the interplay of the adhesion energy $E_{\rm ad}$ of the particles and the bending energy $E_{\rm be}$ of the membranes. The total energy $E$ is the sum 
\begin{equation}
E = E_{\rm be} + E_{\rm ad}
\end{equation}
of these energies. The bending energy of the membrane is the integral
\begin{equation}
E_{\rm be} = 2\kappa \int M^2\, \rm{d}A
\label{Ebe}
\end{equation}
over the area $A$ of the membrane with local mean curvature  $M$ and bending rigidity $\kappa$ \cite{Helfrich73}. We focus here on wrapping scenarios in which the characteristic dimensions of the particles are much smaller than the inverse spontaneous curvature and the characteristic length $\sqrt{\kappa/\sigma}$ of the membranes where $\sigma$ denotes the membrane tension. The spontaneous membrane curvature is then negligible, and the bending energy (\ref{Ebe}) dominates over the membrane tension. The adhesion energy of the particles is
\begin{equation}
E_{\rm ad} = \int \sum_{i=1}^{n_p} V(d_i) dA 
\label{Ead}
\end{equation}
where $V$ is an adhesion potential that depends on the local distance $d_i$ of the membrane from particle $i$, and $n_p$ is the number of particles in contact with the membrane. 

For particles with homogeneous surfaces, the interplay of bending and adhesion energies during wrapping depends on the size and shape of the particles. The shape of a prolate particle can be characterized by its aspect ratio $r = b/a$ where $2 b$ is the distance between the tips of the particles, and $2 a$ is the smallest distance between the sides of the particles at the equator.  We characterize the size of a prolate particle by the radius $R$ of the sphere that has the same surface area as the prolate. The minimum-energy conformations for the wrapping of prolate particles depend on the aspect ratio $r$ and on the rescaled adhesion energy $u = U R^2/\kappa$ where $U$ is the adhesion energy per area, i.e.\ the depth of the adhesion potential $V$.  For patchy particles with inhomogeneous surface segments, the wrapping process depends also on the area fraction of these segments.

%%%
\section{Results} 
%%%

%
\subsection{Prolate particles with adhesion potentials of negligible range} 
\begin{figure}[htp]
\resizebox{\columnwidth}{!}{\includegraphics{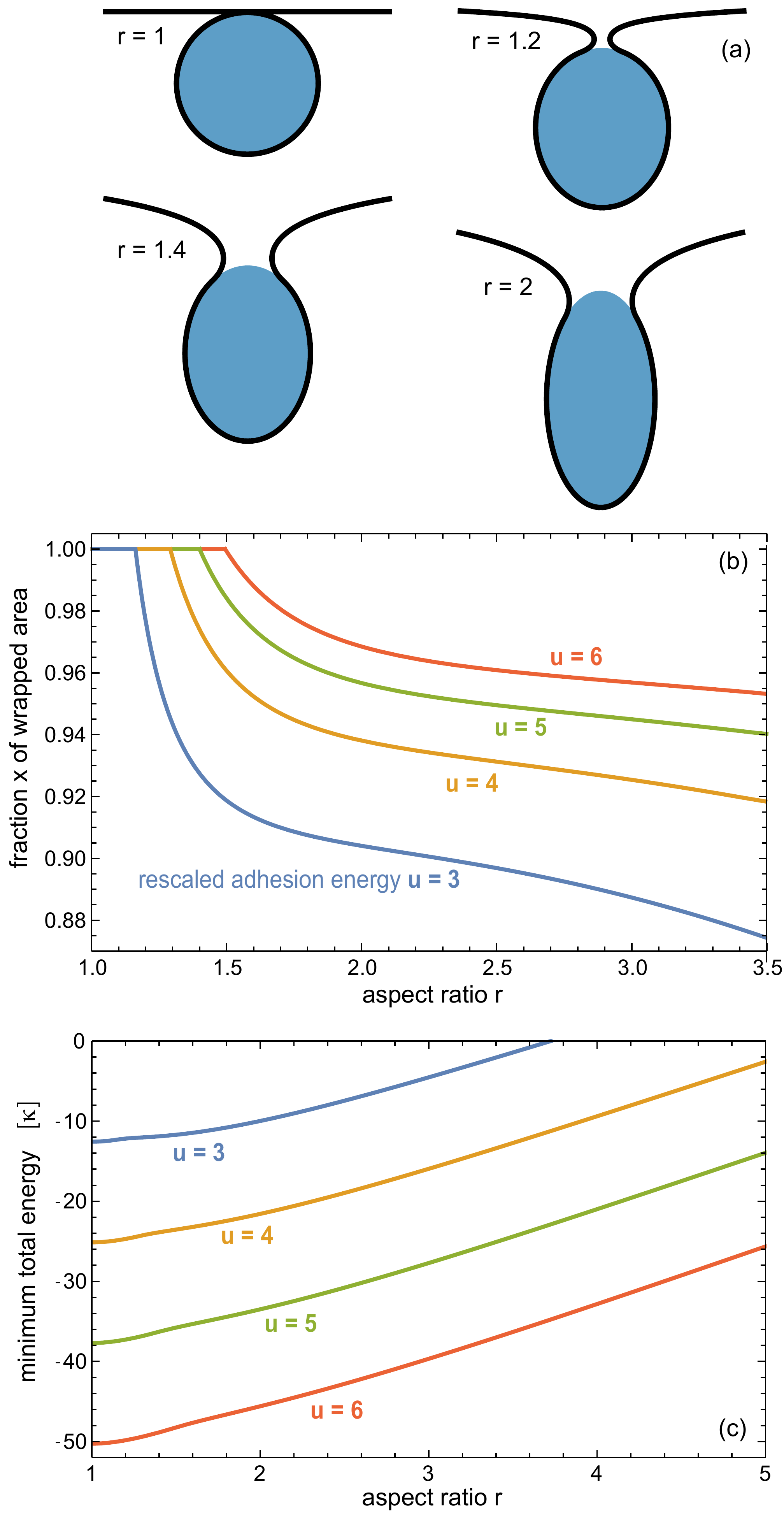}}
\caption{Individual wrapping of prolate particles for negligibly small ranges of the particle-membrane interaction: (a) Minimum-energy membrane profiles for different aspect ratios $r$ of the particles at the rescaled adhesion energy $u=3$. (b) Fraction $x$ of the wrapped particle area and (c) minimum total energy as a function of the aspect ratio $r$ for different rescaled adhesion energies $u$.
}
\label{figure-0-single}
\end{figure}

We first consider the wrapping of single prolate particles by an initially planar membrane. If the range of the particle-membrane interaction is negligibly small compared to the typical dimensions of the particle, the membrane around a particle can be divided into a bound membrane segment that is tightly wrapped around a part of the particle or around the whole particle, and an unbound membrane segment. Deeply wrapped prolate particles are oriented perpendicular to the membrane plane as in Fig.\ \ref{figure-0-single}(a) in their minimum-energy conformation \cite{Bahrami13,Dasgupta14}. In this orientation, the membrane shape around the particles is rotationally symmetric, and the unbound membrane segment adopts a catenoidal shape with bending energy zero for negligible spontaneous curvature $c_o$ and tension $\sigma$ of the membrane. The zero bending energy of the unbound catenoidal segments results from oppositely equal principal curvatures  $c_1=-c_2$, which lead to a mean curvature $M=(c_1 + c_2)/2= 0$. Since the bending energy and adhesion energy of the unbound catenoidal membrane segment are zero, the wrapping of the particle is determined by the interplay of bending and adhesion in the bound membrane segment. 

Fig.\ \ref{figure-0-single} illustrates how the fraction $x$ of the wrapped particle area and the minimum total energy depend on the aspect ratio $r$ for different values of the rescaled adhesion energy $u = U R^2/\kappa$ where U is the adhesion energy per area, $R$ is the radius of a sphere with the same surface area, and $\kappa$ is the bending energy of the membrane.  For a spherical particle with aspect ratio $r=1$, the mean curvature of the bound membrane segment is  $M=1/R$ where $R$ is the radius of the particle, and the bending energy is $E_{\rm be} = 8 \pi \kappa\, x$. The bending energy $E_{\rm be}$ is positive and opposes wrapping, while the adhesion energy $E_{\rm ad}  = - 4 \pi  R^2 U\, x$ favors wrapping. For rescaled adhesion energies $u>2$ as in Fig.\ \ref{figure-0-single}, the total energy $E= E_{\rm be}+ E_{\rm ad} = 4\pi\kappa x(2 - u)$ of a spherical particle with aspect ratio $r = 1$ is minimal in the fully wrapped state with $x = 1$. In this fully wrapped state, the catenoidal membrane neck that connects the wrapped membrane segment to the surrounding planar membrane is infinitesimally small. 

\begin{figure}[htp]
\resizebox{\columnwidth}{!}{\includegraphics{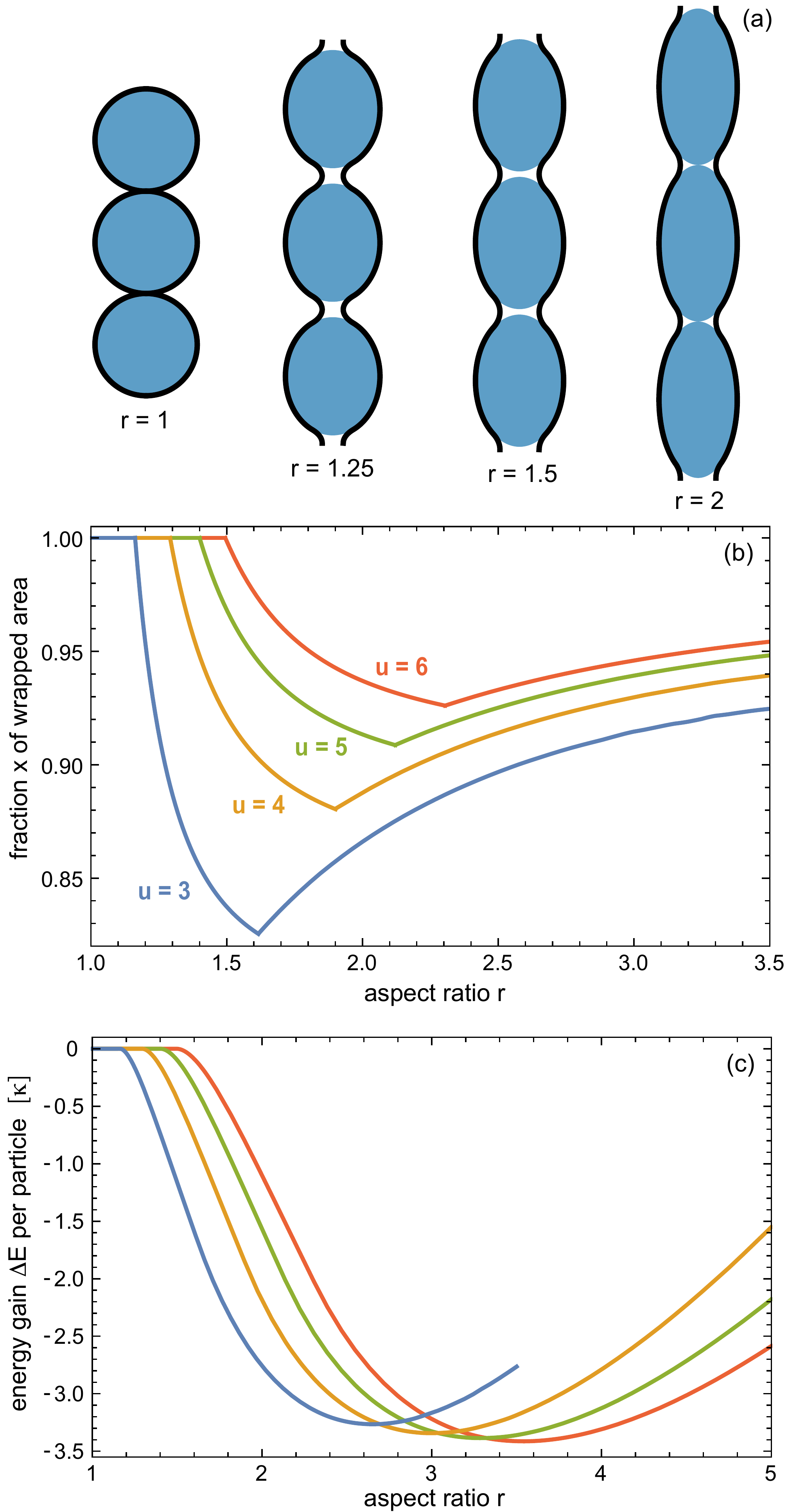}}
\caption{Cooperative wrapping of prolate particles in membrane tubules for negligibly small ranges of the particle-membrane interaction: (a) Minimum-energy membrane profiles around thee central particles in tubules for different aspect ratios $r$ of the particles at the rescaled adhesion energy $u=3$. (b) Fraction $x$ of the wrapped particle area and (c) energy gain $\Delta E$ per particle compared to individual wrapping as a function of the aspect ratio $r$ for different rescaled adhesion energies $u$.
}
\label{figure-0-tube}
\end{figure}

Prolate particles with aspect ratio $r > 1$ have a mean curvature $M$ at their tips that is larger than the mean curvature $1/R$ of a sphere with the same surface area, and a mean curvature at their sides that is smaller than $1/R$. The mean curvature at the tips increases with the aspect ratio $r$. For a given rescaled adhesion energy $u$, prolate particles are fully wrapped for small aspect ratios $r < r_1$ for which the adhesion energy compensates the local bending energy cost of wrapping both at the sides and tips of the particles. For larger aspect ratios $r > r_1$, the prolate particles are partially wrapped.  In the partially wrapped states of single prolates relevant here, one of the tips remains unwrapped, which `safes' bending energy (see Fig.\ \ref{figure-0-single}(a)). The threshold value $r_1$ of the aspect ratio below which prolate particles are fully wrapped increases with the rescaled adhesion energy $u$ and attains the values $r_1 = 1.16$ for $u = 3$, $r_1 = 1.29$ for $u = 4$, $r_1 = 1.40$ for $u = 5$, and $r_1 = 1.49$ for $u = 6$ (see Fig.\ \ref{figure-0-single}(b)). For aspect ratios $r>r_1$, the  fraction $x$ of the wrapped particle area decreases because of the increasing bending energy cost at the tip of the particle. The increase of the minimum total energy with $r$ shown in Fig.\ \ref{figure-0-single}(b) results from an increase of the overall bending energy cost of wrapping for a given rescaled adhesion energy $u$.

How the cooperative wrapping of prolate particles in tubules depends on the aspect ratio $r$ of the particles is illustrated in Fig.\ \ref{figure-0-tube}. For small aspect ratios $r< r_1$, the particles are fully wrapped in the tubules, and neighboring particles are connected by infinitesimally small catenoidal membrane segments (see profile in Fig.\ \ref{figure-0-tube}(a) for $r=1$). The energy gain $\Delta E$ per particle relative to the individual wrapping is 0  for $r< r_1$ because the particles are fully wrapped both in tubules and as single particles, with identical total energies. The energy gain $\Delta E$ is the difference in total energy for a particle in a tubule and a single wrapped particle. For aspect ratios $r>r_1$, the particles are only partially wrapped, and the wrapping in tubules becomes energetically favorable with negative values of $\Delta E$. The energy gain $\Delta E$ for cooperative wrapping results from the fact that both tips are unwrapped in the tubules, which reduces the bending-energy cost of wrapping, compared to individual wrapping for which only one of the tips is unwrapped (see profiles in Fig.\ \ref{figure-0-single}(a)  and \ref{figure-0-tube}(a) for $r >1$). 

The fraction $x$ of the wrapped surface area of prolate particles in tubules is minimal at an aspect ratio $r_2$, which attains the values $r_2 = 1.62$ for $u = 3$, $r_2 = 1.90$ for $u=4$, $r_2 = 2.12$ for $u = 5$, and $r_2 = 2.30$ for $u=6$  (see Fig.\ \ref{figure-0-tube}(b)). The fraction $x$ of the wrapped particle area decreases for aspect ratios $r$ with $r_1 < r < r_2$, and increases with increasing aspect ratio for $r>r_2$. For large aspect ratio $r>r_2$, neighboring particles in the tubes are in direct contact (see profile in Fig.\ \ref{figure-0-tube}(a) for $r=2$), and the unbound membrane segment cannot attain a catenoidal shape of zero bending energy, because the catenoid does not `fit' between the particles. For these aspect ratios, the cooperative wrapping in tubules  is determined by the interplay of the bending energies in both the bound and unbound membrane segments, and the adhesion energy of the bound segment. For aspect ratios $r$ with $r_1 < r < r_2$, in contrast, neighboring particles in the tubules are not in direct contact (see profiles in Fig.\ \ref{figure-0-tube}(a) for $r=1.25$ and $r = 1.5$), and the unbound membrane segments are catenoids.  
 
The energy difference $\Delta E$ per particle between tubular and individual wrapping is minimal at an aspect ratio $r$ larger than $r_2$ (see Fig.\ \ref{figure-0-tube}(c)). At these minima, the cooperative wrapping in tubulus is energetically most favorable compared to the wrapping as single particles. The minimal value of $\Delta E$ slightly decreases with increasing rescaled adhesion energy $u$, from $\Delta E = -3.27 \, \kappa$ for $u = 3$ to $\Delta E = -3.41 \, \kappa$ for $u = 6$. The location of the minimum shifts from $r = 2.65$ for  $u = 3$ to $r = 3.54$ for $u = 6$. For $u = 3$, the energy gain $\Delta E$ is only shown for aspect ratios $r< 3.5$ because the wrapping of single particles in the perpendicular orientation becomes unstable for larger aspect ratios at this rescaled adhesion energy, compared to the unwrapped state with energy 0, or to weakly wrapped states in which individual prolate particles are lying with their sides on the membrane \cite{Bahrami13,Dasgupta14}. For larger rescaled adhesion energies $u\ge 4$, the total energy in the perpendicular orientation is negative for aspect ratios up to $r = 5$ considered here (see Fig.\ \ref{figure-0-single}(c)), and thus lower than the energy zero of the unwrapped state.

%%%
\subsection{Prolate particles with an adhesion potential of finite range $\rho$} 
%%%

For particles with a characteristic dimensions of tens of nanometers, the range $\rho$ of the particle-membrane interaction is typically not negligibly small compared to these particle dimensions. To explore the effect of the potential range on wapping, we consider here the potential
\begin{equation}
V(d)=U\left(\frac{\rho^{9}}{2(\rho+d)^{9}} - \frac{3\, \rho^3}{2(\rho+d)^3}\right) 
\label{potential}
\end{equation}
which adopts its minimum value $-U$ at the relative distance $d=0$. The relative distance $d=0$ thus corresponds to the equilibrium distance between the particle surface and a bound membrane site in the absence of other than adhesive forces. The adhesion potential (\ref{potential}) reflects Lennard-Jones type interactions between a membrane site and the whole particle. The parameter $\rho$ is a characteristic length that determines the range of the attractive interactions in the adhesion potential (\ref{potential}).

\begin{figure}[htb]
\resizebox{\columnwidth}{!}{\includegraphics{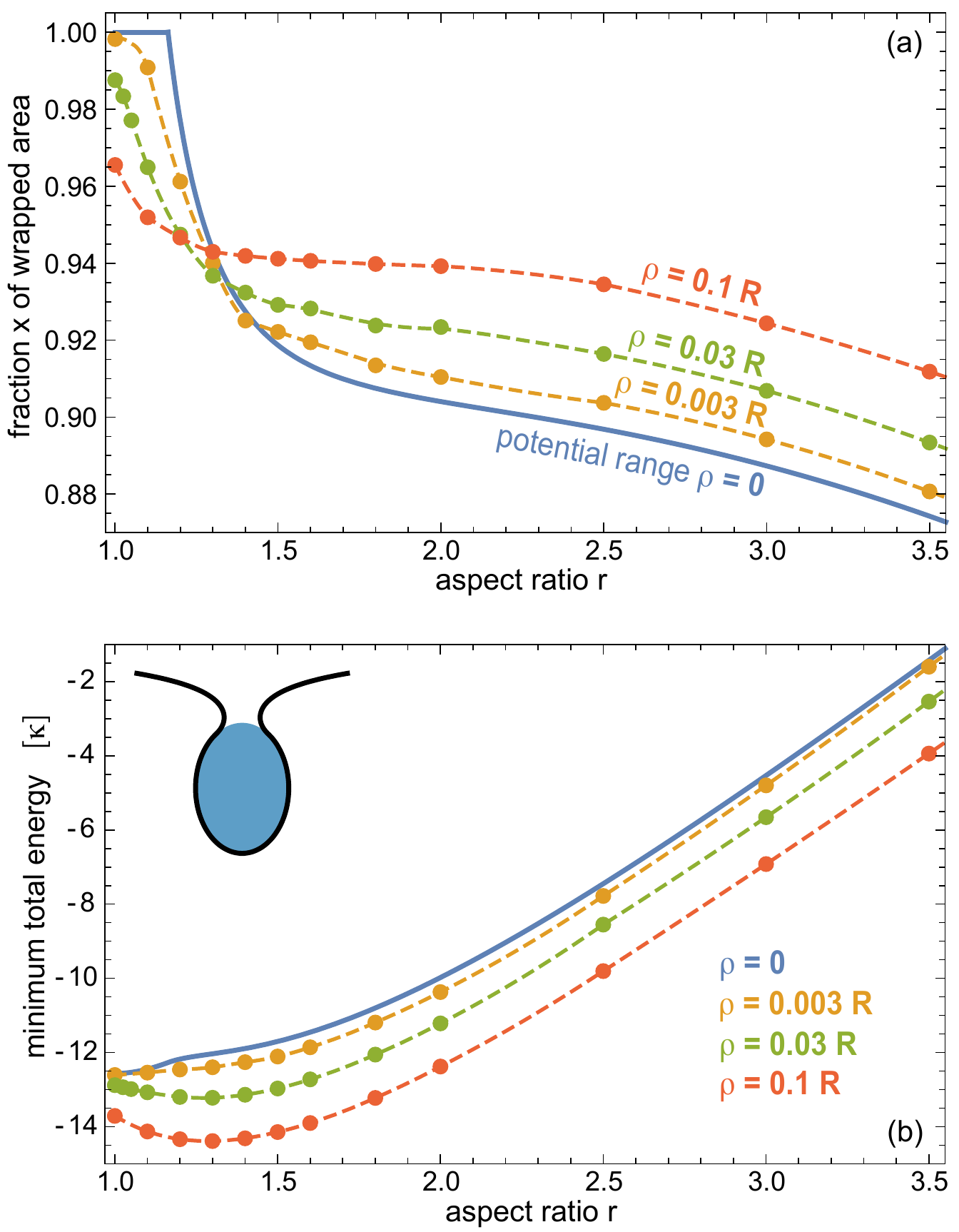}}
\caption{Individual wrapping of prolate particles for the particle-membrane interaction potential (\ref{potential}) with characteristic range $\rho$:  (a) Fraction $x$ of the wrapped particle area and (b) minimum total enery $E$ as a function of the aspect ratio $r$ for the rescaled adhesion energy $u = 3$ and different values of $\rho$. The full blue lines for $\rho= 0$ correspond to the blue lines in Fig.\ \ref{figure-0-single}. The dashed interpolation lines of the data points for finite $\rho$ are guides for the eye.
}
\label{figure-LJ-single}
\end{figure}

\begin{figure}[htb]
\resizebox{\columnwidth}{!}{\includegraphics{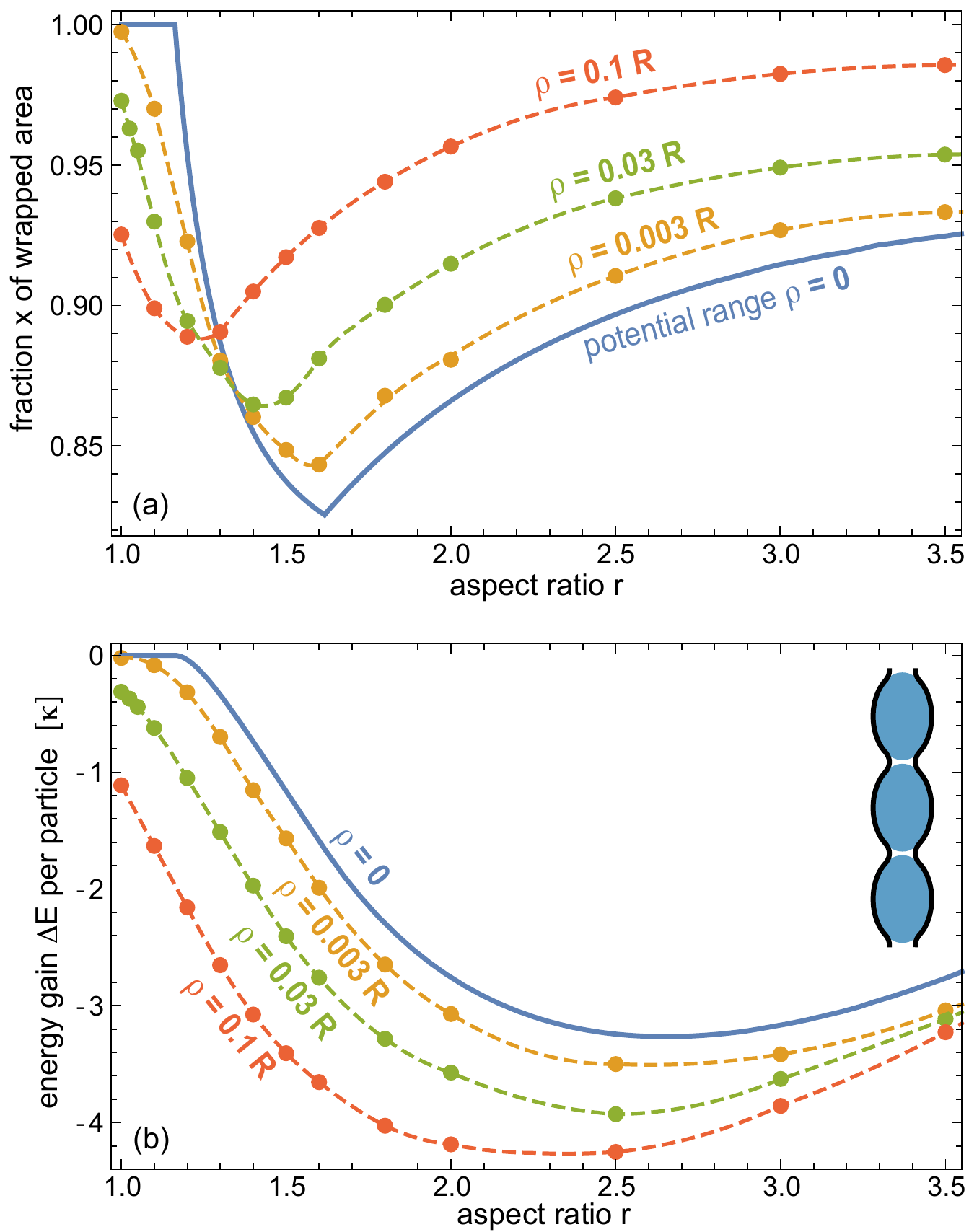}}
\caption{Cooperative wrapping of prolate particles in membrane tubules for the particle-membrane interaction potential (\ref{potential}) with characteristic range $\rho$:  (a) Fraction $x$ of the wrapped particle area and (b) energy gain $\Delta E$ per particle compared to individual wrapping as a function of the aspect ratio $r$ for the rescaled adhesion energy $u = 3$ and different values of $\rho$. The full blue lines for $\rho= 0$ correspond to the blue lines in Fig.\ \ref{figure-0-tube}. The dashed interpolation lines of the data points for finite $\rho$ are guides for the eye.
}
\label{figure-LJ-tube}
\end{figure}

Fig.\ \ref{figure-LJ-single} illustrates how the wrapping of a single prolate particle depends on the potential range $\rho$ and aspect ratio $r$ for the rescaled adhesion energy $u = 3$. The full blue lines are for a negligible small potential range $\rho$ and correspond to the blue lines in Fig.\ \ref{figure-0-single}. For finite potential range $\rho$, prolate particles are partially wrapped at all aspect ratios $r$ (see Fig.\ \ref{figure-LJ-single}(a)). In these partially wrapped states, one of the particle tips remains unwrapped. For small values of $r$ close to 1, the fraction $x$ of the wrapped area decreases with increasing potential range $\rho$. For large values of $r$, the wrapped area increases with $\rho$. The minimum total energy $E$ decreases with increasing potential range $\rho$. This decrease of $E$ results from a favorable interplay of bending and adhesion energies in the boundary region in which the membrane detaches from the particle \cite{Raatz14}. In this boundary region between the bound and unbound membrane segments, the membrane already approaches the catenoidal shape of the unbound membrane segment with zero bending energy, but still gains adhesion energy due to the finite potential range $\rho$. With increasing potential range $\rho$, the minimum energy $E$ decreases because the boundary region between the bound and unbound membrane segments becomes wider \cite{Raatz14}.
Similar to the wrapping of single particles, the fraction $x$ of the wrapped area for particles in tubules decreases with increasing potential range $\rho$ at small values of the aspect ratio $r$ close to 1 (see Fig.\ \ref{figure-LJ-tube}(a)). At large values of $\rho$, the wrapped area fraction $x$ increases with $\rho$. 


The energy difference $\Delta E$ per particle between tubular and individual wrapping decreases with increasing potential range $\rho$ for all aspect ratios $r$ (see Fig.\ \ref{figure-LJ-tube}(b)). This decrease of $\Delta E$ indicates that cooperative wrapping becomes more favorable with increasing potential range, compared to individual wrapping. The energy difference $\Delta E$ decreases with $\rho$ because a particle in the tubule has two boundary regions between bound and unbound membrane segments, since both tips of the particle are unwrapped in the tubule for finite $\rho$. In contrast, a single wrapped particle has only a single boundary region with favorable interplay between bending and adhesion energies. For finite potential range $\rho$,  the energy gain $\Delta E$ is nonzero for all aspect ratios $r$ of prolate particles because the particles then are partially wrapped for all $r$. Cooperative wrapping in tubules is energetically favorable only for partially wrapped particles because the energies of full wrapping are identical for particles in tubules and for single particles. 

%%%
\subsection{Triblock Janus particles}
%%%

The results for prolate particles presented in the previous sections indicate that the cooperative wrapping in tubules is energetically favorable if the particles are partially wrapped in their minimum-energy states. Prolate particles are partially wrapped whenever not wrapping the more highly curved tips `saves energy', and the cooperative wrapping in tubules is then favorable because both tips can remain unwrapped. Cooperative wrapping in tubules can also be favorable for other particles with opposing tips or surface segments that `resist' wrapping. This resistance to wrapping can result either (i) from an increased bending energy cost of wrapping the particle tips, as in the case of prolate particles or other elongated particles with more strongly curved tips, or (ii) from low or vanishing adhesion energies at the opposing tips or surface segments.

\begin{figure}[t]
\resizebox{\columnwidth}{!}{\includegraphics{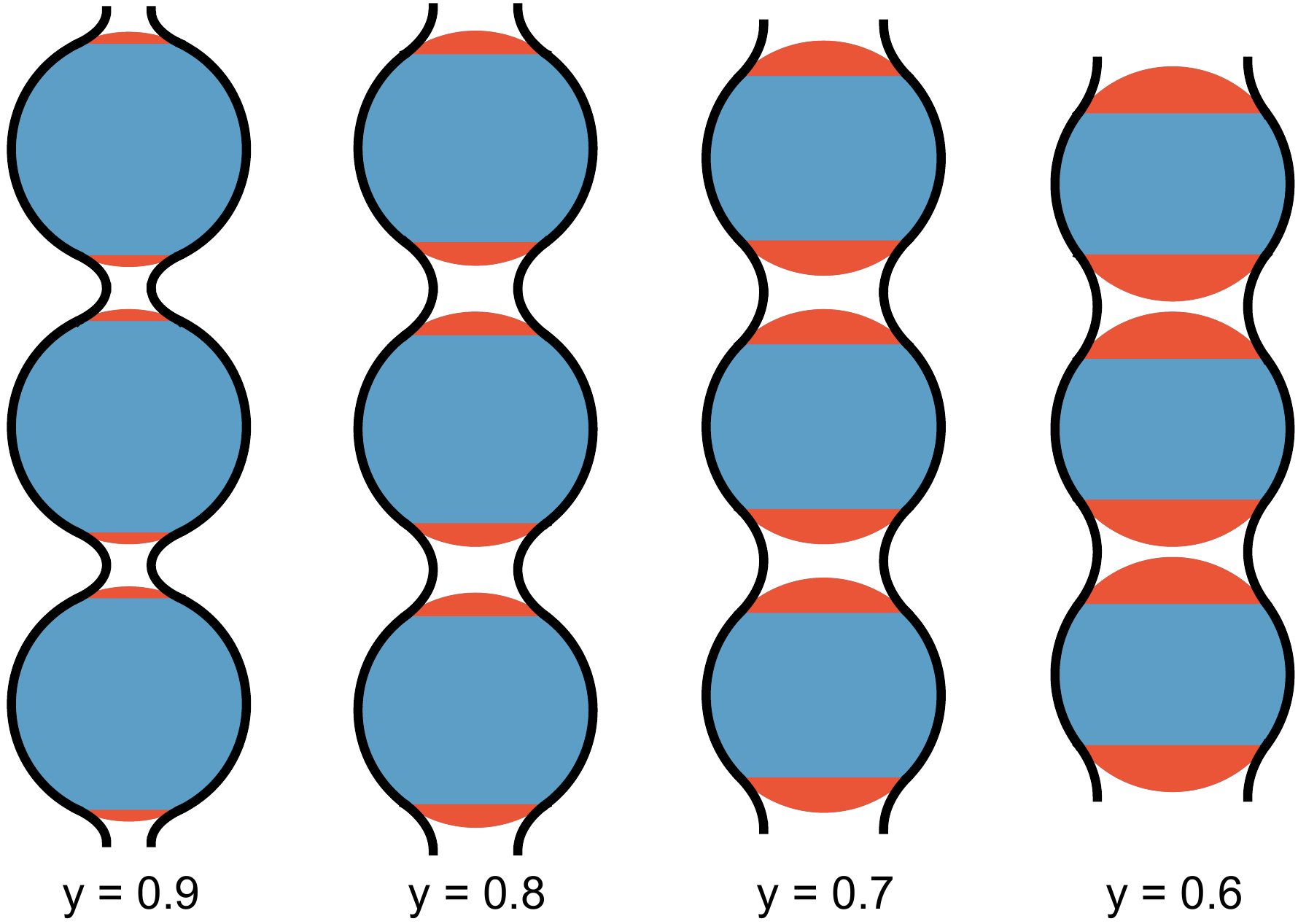}}
\caption{Minimum-energy profiles of cooperatively wrapped triblock Janus particles with weakly adhesive or non-adhesive caps (red) and a rather strongly adhesive central surface segment (blue) for different area fractions $y$ of this segment. In these membrane tubules, the wrapping degree $x$ of the particles is identical to the area fraction $y$ of the adhesive central surface segment. 
}
\label{figure-janus}
\end{figure}

As an example, we consider here spherical triblock Janus particles with opposing weakly adhesive or nonadhesive caps and a more strongly adhesive central surface segment between these caps. Not wrapping the caps saves energy if the rescaled adhesion energy $u_{\rm cap}$ at the caps is smaller than 2, because the  bending energy cost of wrapping the caps then is not compensated by the adhesion energy. In membrane tubules, both caps can remain unwrapped, and the overall wrapping degree $x$ is identical to the area fraction $y$ of the adhesive central surface segment (see Fig.\ \ref{figure-janus}). For individually wrapped triblock Janus particles, the central surface segment is fully covered by the membrane for sufficiently large adhesion energies of this segment. However, the individual wrapping of triblock Janus particle then requires to wrap one of two weakly adhesive caps, with energy cost $2\pi \kappa(1-y)(2-u_{\rm cap})$ (see also section 3.1). This energy cost for wrapping a single cap corresponds to the energy gain for the cooperative wrapping in tubules, at least for area fractions $y$ of the adhesive central segment larger than about $0.57$ as in Fig.\ \ref{figure-janus}. For such area fractions $y$, the non-adhering membrane between neighboring particles adopts a catenoidal shap with zero bending energy. For smaller area fractions of the adhesive central segment, catenoids do not fit between the particles, and the bending energy of the non-adhering membrane has to be taken into account in the calculation of the energy gain of cooperative wrapping.

\section{Discussion and Conclusions}

For prolate particles and triblock Janus particles, the cooperative wrapping in membrane tubules is highly stable relative to individual wrapping for all ranges of the particle-membrane interaction. With increasing interaction range $\rho$, the stability of membrane tubules filled with prolate particles increases. Since typical values of the bending rigidity $\kappa$ of reconstituted lipid membranes \cite{Nagle13,Dimova14} and of cellular membranes \cite{Pontes13,Betz09} range from 10 $k_B T$ to 80 $k_B T$, the energy differences $\Delta E$  between tubular and individual wrapping of prolate particles obtained from energy minimization are  in general large compared to the thermal energy $k_B T$ (see Fig.\ \ref{figure-0-tube}(c)) and \ref{figure-LJ-tube}(b)). For spherical particles with aspect ratio $r = 1$, the energy gain $\Delta E$ for cooperative wrapping strongly depends on the range of the particle-membrane interaction, and vanishes for negligible interaction range because the particles then are fully wrapped both in tubules and as individual particles, with identical minimum energies. 

The high stability of particle-filled membrane tubules implies  strongly attractive elastic interactions between the particles that are mediated by the membrane. In addition to these membrane-mediated interactions, the formation of membrane tubules can be affected by direct interactions between the particles, which are neglected in our model. In experiments, the direct interactions of nanoparticles are typically repulsive or only weakly attractive to prevent aggregation in the bulk solution. However, neighboring particles in tubules need not be in close contact (see Figs.\ \ref{figure-LJ-tube}(a) and \ref{figure-janus}), and repulsive direct interactions  only play a role for tubular stability if their range is comparable to the distance between neighboring particles in the tubes. 

We have focused here on the stability of membrane tubules and, thus, on equilibrium aspects of the cooperative wrapping in tubules. As dynamic processes, the cooperative wrapping in tubules and the individual wrapping of particles are in competition. It seems reasonable to assume that particles with large adhesion energies are first wrapped as single particles as soon as they come in contact with the membrane. However, these particles are only partially wrapped in situations in which membrane tubules are stable, which impedes membrane fission processes of the catenoidal membrane neck that connects the wrapped particle to the surrounding membrane. Such fission processes in general require fully or nearly fully wrapped states in which the apposing membranes of the catenoidal neck are in close contact. In partially wrapped states, individually wrapped particles are likely to remain at the membrane, and membrane tubules may eventually form due to the membrane-mediated interactions between these particles.

{\em Acknowledgements.}---  Financial support from the Deutsche Forschungsgemeinschaft (DFG) {\em via} the International Research Training Group 1524 ``Self-Assembled Soft Matter Nano-Structures at Interfaces" is gratefully acknowledged.

%%%
\section{Methods}
%%%

%%%
\subsection{Energy minimization for adhesion potentials of negligible range} 
%%%

For particle-membrane interactions with a range that is negligibly compared to particle dimensions, bound membrane segments tightly adhere to the particle and, thus, have the same shape as the particle. The shape of a prolate particle can be described by
\begin{equation}
(\phi,\psi) \rightarrow \vec{r}(\phi,\psi)= \left( \begin{array}{c} a\, \sin \psi \, \cos \phi \\ a\, \sin \psi \, \sin \phi \\ b\ \cos \psi \\ \end{array}
\right)
\label{eq:Para1}
\end{equation}
with $b>a$, $0\le \phi < 2 \pi$, and $0\le \psi\le \pi$. The aspect ratio of the prolate is $r = b/a$. A rotationally symmetric, bound membrane segment has the bending energy density 
\begin{equation}
e_{\rm be}(\psi) =\kappa \frac{\sqrt{2}\pi r^2 \left(3+r^2 - (r^2-1) \cos 2 \psi \right)^2}
   {\left(1 + r^2 - (r^2-1) \cos 2 \psi\right)^{5/2}}
\end{equation}
and bending energy $E_{\rm be} = \int e_{\rm be}(\psi) \sin \psi \,\mathrm{d}\psi$, and the adhesion energy density
\begin{equation}
e_{\rm ad}(\psi) =- U R^2 \frac{\sqrt{8}\pi \sqrt{r^4 - 1 - (r^2-1)^2 \cos 2 \psi}}
 {\sqrt{r^2 - 1} + r^2 \, \mathrm{arcsec}\, r}
\end{equation}
and adhesion energy $E_{\rm ad} = \int e_{\rm ad}(\psi) \sin \psi \,\mathrm{d}\psi$, where $R$ is the radius of a sphere with the same area as the prolate. A prolate particle is fully wrapped by the membrane if the total energy density $e(\psi) = e_{\rm be}(\psi) + e_{\rm ad}(\psi)$ is negative for all $\psi$. The particle is partially wrapped if $e(\psi)$ is negative at the sides and positive at the tips. For the individually wrapped prolate particles of Fig.\ \ref{figure-0-single}, we determine the boundary point $\psi_b$ between the bound and unbound segments  from the condition $e(\psi_b) = 0$. For this boundary point, the total energy of an individually wrapped particle is minimal because the unbound membrane segment adopts a catenoidal shape of zero bending energy. For cooperatively wrapped prolate particles of Fig.\ \ref{figure-0-tube} with aspects ratios $r_1< r < r_2$, the total energy is minimal for the two boundary points $\psi_b$ and $\pi - \psi_b$ at which the total energy density $e(\psi)$ is zero. Particles with these aspect ratios are partially wrapped, and neighboring particles in the tubules are connected by catenoidal membrane segments of zero bending energy. For prolate particles with aspect ratios $r> r_2$, we use the discrete energy minimization in cylindrical coordinates described in the next section to determine the shape of the unbound membrane segment. 

\subsection{Energy minimization for adhesion potentials with finite range}

For the particle-membrane interaction (\ref{potential}) with a finite range $\rho$, we use a discrete energy minimization to determine the profiles of the rotationally symmetric membrane shapes around a single prolate particle and around linear aggregates of particles in tubules. We use both spherical and cylindrical coordinates to describe the rotationally  symmetric membrane shapes. 

In {\em spherical coordinates}, the membrane shapes can be described by
\begin{equation}
(\phi,\psi) \rightarrow \vec{r}(\phi,\psi)= \left( \begin{array}{c} p(\psi)\, \sin \psi \, \cos \phi \\ p(\psi)\, \sin \psi \, \sin \phi \\ p(\psi)\ \cos \psi \\ \end{array}
\right)
\label{eq:Para1}
\end{equation}
with membrane profile $p(\psi) \geq 0$, $0\le \phi < 2 \pi$, and $0\le \psi\le \pi$. In this parametrization, the bending energy (\ref{Ebe}) and adhesion energy (\ref{Ead}) adopt the form 
\begin{equation}
E_{be}=\!\!\int \!\!\pi \kappa \frac{ \left( p (p p'' - 2p^2 - 3 p'^2) + p'(p'^2 + p^2) \cot\psi\right)^2\!\!}
{p\left(p'^2+p^2\right)^2 \sqrt{p'^2+p^2}} \sin \psi\, \mathrm{d}\psi
\end{equation}
and
\begin{equation}
E_{ad} = 2\pi \int \sum_{i=1}^{n_p} V(d_i) p \sin \psi \sqrt{p^2 + p'^2}  \mathrm{d}\psi
\end{equation}
with $p = p(\psi)$, $p' = \mathrm{d} p(\psi)/\mathrm{d}\psi$, and $p'' = \mathrm{d}^2 p(\psi)/\mathrm{d}\psi^2$, where $d_i = d_i(\psi,p(\psi))$ is the shortest distance of a membrane patch located at $\psi$ and the surface of particle $i$. For a prolate particle, this shortest distance is in general not on the line that connects the membrane patch and the particle center, and has to be determined numerically.  

In {\em cylindrical coordinates}, the membrane profiles are described by the local radial distance $r(z)$ at position $z$ along the axis of rotation:
\begin{equation}
\vec{r}(z,\phi) = \left(
\begin{array}{c}
r(z)\cos\phi\\
r(z)\sin\phi\\
z
\end{array}
\right)
\end{equation}
In this parametrization, the bending energy (\ref{Ebe}) and adhesion energy (\ref{Ead}) adopt the form 
\begin{equation}
E_{\rm be}=\pi \kappa\int \frac{\left(r(z)r^{\prime\prime}(z)-r^\prime(z)^2-1\right)^2}{r(z)\left(r^\prime(z)^2+1\right)^{5/2}}{\rm d}z
\label{Ebe_1}
\end{equation}
and
\begin{equation}
E_{\rm ad} = 2\pi\int  \sum_{i=1}^{n_p} V(d_i)r(z)\sqrt{1+r^\prime(z)^2}{\rm d}z
\label{Ead_1}
\end{equation}
with $d_i = d_i(z,r(z))$. The primes here indicate derivatives with respect to $z$. 

We use cylindrical coordinates to describe the shape of membrane tubules. Because of the periodicity of the membrane tubules and symmetry of the particles, we only consider 
a tubular membrane segment for half a period of the membrane profile in our numerical minimization, i.e.\ a tubular membrane segment from a maximum of the radial distance $r$ 
to the next minimum of $r$. In this minimization, we take into account the adhesion energy of the two particles adjacent to the minimum of $r$.
To describe the rotationally membrane shapes around a single prolate, we use spherical coordinates for the bound membrane segment that is wrapped around one half of the particle, and cylindrical coordinates for the remaining bound and unbound membrane segments. We discretize the profile functions $p(\psi)$ and $r(z)$ of the two parametrizations using up to 1000 discretization points and express the first and second derivatives of these functions as finite differences. The discretization increments $\Delta \psi$ and $\Delta z$ for a single prolate are matched by $\Delta z=a  \Delta \psi$ where $a$ is the radius of the particle's equator at which we switch the parametrizations. We obtain the minimum-energy shapes then from a constrained minimization of the total energy with respect to the values $p(\psi)$ and $r(z)$ at the discretization points using the program Mathematica \cite{Mathematica}.

\end{document}